%% file: main.tex
\documentclass[conference]{IEEEtran}
\IEEEoverridecommandlockouts
\usepackage{cite}
\usepackage{amsmath,amssymb,amsfonts}
\usepackage{algorithmic}
\usepackage{graphicx}
\usepackage{textcomp}
\usepackage{xcolor}
\newtheorem{theorem}{Theorem}

\newtheorem{definition}{Definition}
\usepackage[T1]{fontenc} 
\usepackage[utf8]{inputenc} 

\def\BibTeX{{\rm B\kern-.05em{\sc i\kern-.025em b}\kern-.08em
    T\kern-.1667em\lower.7ex\hbox{E}\kern-.125emX}}
\begin{document}

\title{
Tuning Quantum Computing Privacy through Quantum Error Correction
}

\author{
\IEEEauthorblockN{Hui Zhong$^{1}$, Keyi Ju$^{2}$, Manojna Sistla$^{1}$, Xinyue Zhang$^{3}$, Xiaoqi Qin$^{2}$, Xin Fu$^{1}$,  Miao Pan$^{1}$}

\IEEEauthorblockA{$^{1}$Department of Electrical and Computer Engineering, University of Houston, Houston, TX, 77204}
\IEEEauthorblockA{$^{2}$State Key Laboratory of Networking and Switching Technology,\\ Beijing University of Posts and Telecommunications, Beijing, China, 100876}
\IEEEauthorblockA{$^{3}$Department of Computer Science, Kennesaw State University, Marietta, GA, 30060}
}


\maketitle

\begin{abstract}
Quantum computing is a promising paradigm for efficiently solving large and high-complexity problems. To protect quantum computing privacy, pioneering research efforts proposed to redefine differential privacy (DP) in quantum computing, i.e., quantum differential privacy (QDP), and harvest inherent noises generated by quantum computing to implement QDP. However, such an implementation approach is limited by the amount of inherent noises, which makes the privacy budget of the QDP mechanism fixed and uncontrollable. To address this issue, in this paper, we propose to leverage quantum error correction (QEC) techniques to reduce quantum computing errors, while tuning the privacy protection levels in QDP. In short, we gradually decrease the quantum noise error rate by deciding whether to apply QEC operations on the gate in a multiple single qubit gates circuit. We have derived a new calculation formula for the general error rate and corresponding privacy budgets after QEC operation. Then, we expand to achieve further noise reduction using multi-level concatenated QEC operation. Through extensive numerical simulations, we demonstrate that QEC is a feasible way to regulate the degree of privacy protection in quantum computing.

\end{abstract}


\section{Introduction }
Quantum computing is a new paradigm that is promising to significantly accelerate the processing of large datasets and complex tasks, compared with traditional computing~\cite{gyongyosi2019survey}.
Such an advance in computing will affect a multitude of research domains such as cryptography \cite{mavroeidis2018impact}, huge optimization problems \cite{ajagekar2019quantum}, and complex climate simulations~\cite{berger2021quantum}. However, similar to classical computing, quantum computing is vulnerable to data privacy leakage, because quantum computers face a variety of threats from internal and external attacks \cite{furrer2012continuous}, such as coherent attacks, entangling attacks, quantum-side-channel attacks, inference attacks, etc. Therefore, how to preserve quantum computing privacy poses great challenges.


In traditional computing, differential privacy (DP) is a powerful privacy protection method. In simple terms, DP protects data by manually adding noises to ensure that the presence or absence of a single data has a negligible effect on any query results. DP seeks a balance between data usability and privacy and has applications in diverse fields, such as data analytics \cite{lin2016differential}, medical and health research \cite{sun2019differential}, intelligent transportation \cite{kargl2013differential}, etc. DP has strict mathematical formulas and usually uses a privacy budget $\epsilon$ to indicate the degree of privacy protection \cite{dwork2006differential}. Recently, DP has been introduced to the quantum domain, and quantum differential privacy (QDP), a novel data protection method, brings an overhaul to quantum computing \cite{hirche2023quantum,watkins2023quantum,du2022quantum}. Existing works have already found that Noisy Intermediate-Scale Quantum (NISQ) devices generate unavoidable noises during quantum computation and can act as potential noise sources for QDP \cite{zhou2017differential, guan2023detecting, hirche2023quantum}. Since the inherent noises are determined through specific factors such as environment or hardware device parameters, they are uncontrollable and fixed. This limitation confines these works to a specific privacy budget, which prevents adjusting it to the target DP budget. Therefore, it is necessary to explore suitable approaches that can change the amount of noises to satisfy the different DP requirements in quantum computing. Generally speaking, when the required amount of noise exceeds the inherent noise generated by quantum computing, additional noises should be added to satisfy the privacy budget. Conversely, when the QDP noise requirement is beneath the inherent noise level, designated methods are in need to reduce the inherent noises.


As a promising solution, quantum error correction (QEC) techniques represent quantum bit information as a high dimensional redundancy quantum bits and target error correction and noise reduction caused by environmental disturbances, quantum operations, etc. QEC can be implemented through different error syndromes and specialized quantum circuits, e.g., Steane codes, Shor codes \cite{roffe2019quantum}, etc. By effectively reducing the noises through QEC, we may execute a more reliable quantum computing process \cite{aliferis2007subsystem, duncan2013verifying, fowler2010surface}.


In this paper, we have keenly observed that the QEC method can address the issue of excessive inherent noises in order to achieve an adjustable DP. Based on the formula between privacy budgets and depolarizing noise in \cite{zhou2017differential}, we gradually tune the inherent noises and adjust the degree of privacy protection, by varying whether QEC operation is applied to single qubit gates or not. We conduct a rigorous mathematical analysis on a general formula for the total error rate after applying the QEC operation and the corresponding privacy budget calculation in a single qubit circuit with multiple gates. We also find that multi-level concatenated QEC can further lower the error rate at the expense of circuit complexity. We demonstrate our ideas through extensive numerical simulations.


The rest of the paper is organized as follows. In Section ~\ref{sec:preliminary}, we introduce the theoretical foundations of QDP, QEC, and the basic privacy budget formulas. In Section ~\ref{sec:QEC}, we present generalized total noise error rate formulas and privacy budget formulas under multiple single qubit gates, and extend to the multi-level concatenated QEC operation. In Section ~\ref{sec:performance}, we provide numerical simulation. Finally, we draw conclusions and discuss future work.

\section{Preliminaries \label{sec:preliminary}}
In this section, we review the basics of quantum computing, traditional DP, and the existing formulation of QDP. Subsequently, we demonstrate how to calculate the privacy budget with depolarizing noise, as proposed in \cite{zhou2017differential}. Finally, we introduce QEC, a method for adjusting the depolarizing noise error rate.
\vspace{-2mm}
\subsection{Quantum computing}
The basic unit used in quantum computing is the qubit, represented by the two states of the standard orthogonal basis $\left | 0  \right \rangle$  and $\left | 1  \right \rangle$, corresponding to the values 0 and 1 in classical computers. A single-qubit can be represented as $\left | 0  \right \rangle = (1,0)^{T}\ $ and $\left | 1  \right \rangle = (0,1)^{T}$. Quantum states describe the states of a quantum system. For example, in a two-dimensional Hilbert space, a pure quantum state can be represented as the wave function of a two-state system, consisting of a superposition of $\left | 0  \right \rangle$  and $\left | 1  \right \rangle$: $\left | \psi  \right \rangle = \alpha_{1}  \left | 0  \right \rangle + \alpha_{2} \left | 1 \right \rangle = (\alpha_{1},\alpha_{2} )^{T}\in \mathbb{C} ^{2} $, where the complex numbers $\alpha_{1}$ and $\alpha_{2}$ satisfy the formula $\left | \alpha_{1}  \right | ^{2} + \left | \alpha_{2} \right | ^{2} = 1$. In D-dimensional Hilbert space, it can be generalized to  $\left | \psi  \right \rangle = \sum_{i=1}^{D}\alpha _{i}\left | i  \right \rangle  \in \mathbb{C} ^{D}$, which satisfies $\left | \alpha_{1}  \right | ^{2} +...+ \left | \alpha_{D} \right | ^{2} = 1$. $\left | i  \right \rangle$ is the ground state, $\alpha _{i}$ is the complex amplitude.

Another form of the quantum state is a mixed state $\rho$, defined as $\rho =\sum_{i=1}^{d} p_{i} \left | \psi  \right \rangle \left\langle\psi\right| $, which can be represented by the density matrix and satisfies $\text{Tr}\left ( \rho  \right ) =1$. The difference between two quantum states can be represented by the trace distance $d$, which can be calculated by the trace of these two density matrices. For quantum states $\rho$ and $\sigma$, their trace distance is denoted as $\left | \left | \rho -\sigma  \right |  \right |  ^{*} =\text{Tr}\left ( \left | \rho -\sigma\right |  \right ) /2$.


Quantum computing is presented by quantum circuits, which consist of multiple types of gates. The quantum gate operations are usually represented by the unitary operator $U$. For example, a single quantum gate can be represented as $2\times 2$ unitary matrix. If a quantum state $\rho$ goes through the $U$ operation, its output can be represented as $\rho'=U\rho U^{\dagger}$, which satisfies $U^{\dagger} U = U U^{\dagger} = I$. $U^{\dagger}$ is the adjoint of $U$ and $I$ is the identity operator. Quantum measurements $\left \{ M_{k}  \right \} _{k\in O } $ get different results for circuits, where $O$ is a finite set of measurement results. If the result of the circuit before the quantum measurement is $\rho$, the probability that the measurement result is $k$ can be expressed as $p_{k} =Tr\left \{ M_{k} \rho \right \} $.

\vspace{-2mm}
\subsection{From classical DP to QDP}
DP can maintain the statistical information of the data while preventing privacy leakage. It ensures that the contribution of any single data sample has a very small impact on the final statistical output. Formal DP and QDP are defined as follows~\cite{zhou2017differential}.

\begin{definition}[\bf{Classical Differential Privacy}] A randomized function $K$ satisfies $(\epsilon,\delta)$-differential privacy if the data sets $D$ and $D^{'}$ differ by only one participant, and all set of outcomes $\mathcal{S} \subseteq Range(\mathcal{M})$ satisfy
    \begin{equation}
		\rm{Pr}\left[\mathcal{K}(D)\in \mathcal{S}\right] \leq e^\epsilon \cdot \rm{Pr}\left[\mathcal{K}(D')\in \mathcal{S}\right]  + \delta,
    \end{equation}
where $\epsilon$ is the privacy budget and $\delta$ is the broken probability.
    \label{dp}
\end{definition}
\vspace{-4mm}
\begin{definition}[\bf{Quantum Differential Privacy}] Given two quantum datasets $\rho$ and $\sigma$ with $\tau(\rho,\sigma)\leq d$, where $d \in (0,1]$ and $\tau(\cdot,\cdot)$ indicates the trace distance. A quantum operation $\mathcal{E} $ is $(\epsilon,\delta)$-differentially private if every $POVM$  (Positive Operator-Valued Measure) $M=\left \{ M_{m}  \right \} $ and all $\mathcal{S} \subseteq Range(\mathcal{M})$ satisfy,
 \begin{equation}
        Pr\left[\mathcal{E} \left ( \rho \right ) \in _{M}  S\right] \leq e^\epsilon \cdot Pr\left[\mathcal{E} \left ( \sigma \right ) \in _{M}  S\right] + \delta.
 \end{equation}

\end{definition}

\subsection{QDP based on depolarizing noise}\label{AA}
During quantum computation, NISQ devices suffer from unavoidable errors such as amplitude noise and bit-flip noise. Although these noises cannot be quantified in simulations, developers have found that depolarizing noise can represent the global noises in a circuit. This is because depolarizing noise is a good approximation of the device error especially when the quantum circuit is deep enough\cite{urbanek2021mitigating}. Depolarizing noise exists in any quantum simulator.

In the presence of depolarizing noise, the input quantum state collapses to a fully mixed state with a certain probability. The depolarizing noise can be expressed in terms of quantum operations as
\vspace{-4mm}
\begin{equation}
    \mathcal{E}_{\mathrm{Dep}}(\rho) = \frac{p I}{D} + (1-p)\rho,
\end{equation}
where $\rho$ is the initial quantum state, $I$ is the unit operator, $p$ is the depolarizing noise error rate, and $D$ is the dimension of the Hilbert space. In the depolarizing channel, the input quantum state is held constant with probability $1-p$ and evolves to any of the possible quantum states with probability $p/D$.

We followed \textit{Theorem 3} from \cite{zhou2017differential} to calculate privacy budgets $\epsilon$ under the effect of depolarizing noise:

\begin{theorem}[\bf{DP with depolarizing noise}]
For all inputs $\rho$ and $\sigma$ satisfy $\tau \left (\rho,\sigma \right ) \le d$, the depolarizing operation $\mathcal{E}_{\mathrm{Dep}}(\mathcal{E}(\rho)) $ in the D-dimension Hilbert space provides $\epsilon$-differential privacy where
\vspace{-1.5mm}
\begin{equation}
\epsilon=\ln_{}{\left [ 1+\frac{1-p}{p}dD \right ] } .
\label{oridp}
\end{equation}
\end{theorem}
\vspace{-1.5mm}

Equation (\ref{oridp}) shows that the privacy budget $\epsilon$ is only related to the error rate $p$ of the depolarizing noise because the other parameters are fixed. Therefore we would like to find some specific ways to reduce $p$ of depolarizing noise in order to decrease inherent noises and further obtain the tuning DP.

\vspace{-2mm}
\subsection{Quantum error correction}
Quantum noise is a catastrophic obstacle in the development of quantum computing. Since qubits are not replicable, classical error correction methods cannot work well at the quantum level. Quantum error correction (QEC), first proposed by Peter Shor in 1994 \cite{shor1994algorithms}, brought a breakthrough in this problem. QEC formulates an error-correction circuit by storing a quantum bit of information in an entangled state with a high number of quantum bits, which can localize and correct quantum bit errors in this high-dimensional Hilbert subspace \cite{roffe2019quantum}. Common QEC methods are Shor code \cite{aliferis2007subsystem}, Steane code \cite{duncan2013verifying}, Surface code \cite{fowler2010surface}, and so on. Different coding methods correspond to different circuits and thus correct different kinds of errors. In this paper, we choose Steane code to correct circuit errors.

\begin{figure}[t]
\centering
\includegraphics[width=0.4\textwidth]{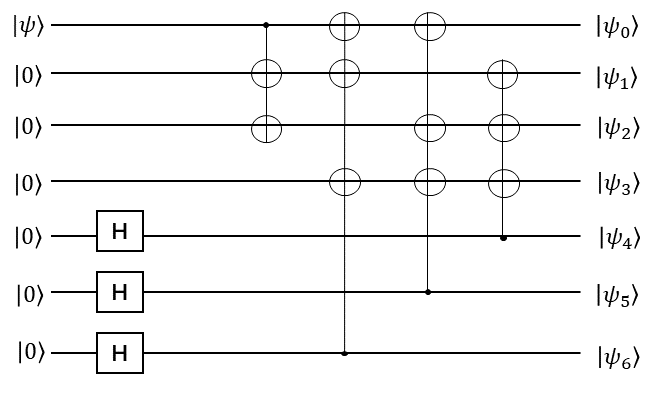}
\caption{Encoding circuit of Steane code.\label{fig:encode}}
\vspace{-4mm}
\end{figure}

\begin{figure}[t]
\centering
\includegraphics[width=0.48\textwidth]{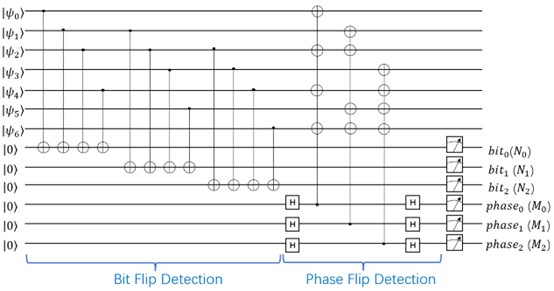}
\caption{Detection circuit of Steane code.\label{fig:detect}}
\vspace{-4mm}
\end{figure}

Steane code corrects any kind of error on a single physical qubit by encoding seven physical bits into one logical qubit. It consists of four modular circuits: encoding, detection, correction, and decoding. The function of the encoding circuit is to encode one logical qubit into seven physical qubits as shown in Fig. \ref{fig:encode}. The purpose is that any single physical qubit error can be detected and corrected by the parity code. The purpose of the detection circuit is to locate which type of error occurs in which physical qubit. The circuit is shown in Fig. \ref{fig:detect}, where the first half is designed to detect bit-flip errors and the second half is designed to detect phase-flip errors, and the combination of the two parts can detect both errors. After detecting the circuit, the auxiliary bits need to be measured to get the six-bit error syndrome. The error syndrome of the Steane code is shown in Table \ref{table*}. As an example, if the measurement result is $\left ( N_{0}, N_{1}, N_{2}, M_{0}, M_{1}, M_{2}\right ) =\left ( -1,1,1,1,1,1 \right ) $, it means that $X$ error occurred in the $0^{th} $ physical qubit. The error syndrome is very important for the error correction circuit. When a specific error at a specific location is obtained from the detect circuit, the error is corrected using the same error type. For example, if an $X$ error occurs in the $0^{th} $ physical qubit, an $X$ gate is added to the $0^{th} $ physical qubit in the error correction circuit to correct the error. Finally, the decoding circuit completes the conversion from seven physical qubits to one logical qubit \cite{quan2022implementation}. 

\input{table.tex}

\section{The impact of QEC on privacy budgets \label{sec:QEC}}
From the previous sections, we can observe that the only parameter impacting $\epsilon$ in Eq. (\ref{oridp}) is the error probability $p$ of the depolarizing noise. In the case of quantum computing with only a single qubit gate, we derive the error probability $p'$ of the depolarizing noise after the QEC. We then extend it to multiple single qubit gates and obtain different error probabilities $p'$ by deciding the location of the QEC, which further satisfies the adjustable privacy budget $\epsilon$ in QDP. Finally, we consider the impact of multi-level concatenated QEC on the QDP. We leave the analysis of $\epsilon$ for multiple qubit gates in future work.

\subsection{DP for a single qubit gate}
First, we continue the setup in \cite{zhou2017differential} by assuming that the quantum circuit only consists of a single qubit gate and replacing the general noises with the depolarizing noise on this gate with an error probability of $p$.

We first encode this single logical qubit into seven physical qubits for subsequent QEC operations. According to the theorem of Steane code \cite{roffe2019quantum}, the depolarizing noise error rate on each physical qubit is $p$. Therefore, when QEC is not applied, the accuracy of one logical qubit is $1-p$; the total accuracy of seven physical qubits is $(1-p)^7$. The Steane code can correct arbitrary errors on a single physical qubit, but it cannot correct the case where multiple physical qubits produce errors at the same time. Thus, it can correct $7\times p\times (1-p)^6$ error rate for seven physical bits. Overall, after QEC operation is applied, a single qubit gate can have the following accuracy and error rate: 
\vspace{-3mm}
\begin{align}
    c'=(1-p)^7+7\times p\times (1-p)^6,
    \label{pstar}
\end{align}
\vspace{-4mm}
\begin{align}
    p'=1-\left[(1-p)^7+7\times p\times (1-p)^6\right],
    \label{pstarr}
\end{align}

\noindent where $c'$ and $p'$ represent the circuit accuracy and the final depolarizing noise error rate respectively.

We substitute $p'$ into the Eq. (\ref{oridp}), and can obtain new formulas for the depolarizing noise error rate and privacy budgets: $\epsilon=\ln_{}{[1+\frac{1-p'}{p'}dD ]}$.


Usually, we hope that QEC can reduce the noise error probability, so the accuracy after using QEC operation needs to be higher than the previous one, which satisfies
\begin{align}
    1-\left[(1-p)^7+7\times p\times (1-p)^6\right]>1-p.
    \label{constraints}
\end{align}
When $p< \approx 0.0579$, QEC operation can reduce the error rate $p$ of the depolarizing noise which can show its own advantage. Therefore, in the simulation, we set the error rate of depolarizing noise between 0 and 0.05.



\subsection{DP for two qubit gates}
In this section, we first use two single qubit gates as an example, assuming that the depolarizing noise as global noises with error rate $p$ occurs on each gate. Without QEC operation, the final accuracy of the logical qubit is $(1-p)\times(1-p)$. 

\begin{figure}[t]
\centering
\includegraphics[width=0.4\textwidth]{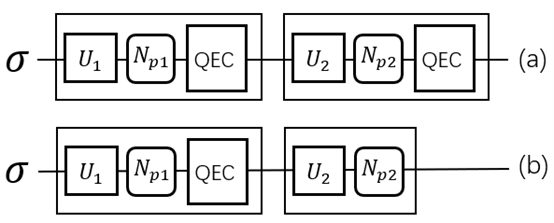}
\caption{(a) QEC operation for each gate. (b) Selective QEC operation for gates. $U_{1}$ and $U_{2}$ are arbitrary single-bit circuit gates. $N_{P1} $ and $N_{P2} $ are depolarizing noise added to the gate. QEC is Steane code circuit.\label{fig:onetwogate}}
\vspace{-4mm}
\end{figure}

We still start by encoding a single logical qubit into seven physical qubits. We propose the following two cases based on whether or not we perform QEC operation for each gate. The circuit schematic is shown in Fig. \ref{fig:onetwogate}.

\textbf{QEC operation for each gate:} In this case, each qubit gate is followed by the QEC operation. According to \cite{du2021quantum}, assuming that the two qubit gates are independent of each other, their final error rate is
\begin{equation}
\begin{split}
    &p_{1}'=1-c'\times c'=1-\left[(1-p)^7+7\times p\times (1-p)^6\right]^2.
\end{split}
\end{equation}
Therefore, Eq. (\ref{oridp}) can evolve into $\epsilon=\ln_{}{[1+\frac{1-p_{1}'}{p_{1}'}dD ]}$.

\textbf{Selective QEC operation for gates:}
In the second case, we selectively perform QEC operations on quantum gates. For a two qubit gates circuit, we perform QEC operation on only one gate. In the case of QEC on the first gate, the depolarizing noise rate consists of the accuracy $c'$ of the first gate after it has been corrected by the QEC and the original accuracy $1-p$ of the second gate. Thus, the total error rate of the depolarizing noise can be expressed as
\begin{equation}
\begin{split}
    &p_{2}'=1-c'\times (1-p)\\
    &=1-\left[(1-p)^7+7\times p\times (1-p)^6\right]\times (1-p) .
\end{split}
\end{equation}
Therefore, Eq. (\ref{oridp}) can evolve into $\epsilon=\ln_{}{[1+\frac{1-p_{2}'}{p_{2}'}dD ]}$.

It is worth noting that since Steane code can only correct errors in one physical qubit and cannot correct errors in two physical qubits, the error rate of the quantum circuit is the same regardless of whether the QEC is located after the first gate or after the second gate.

\begin{figure}[t]
\centering
\includegraphics[width=0.45\textwidth]{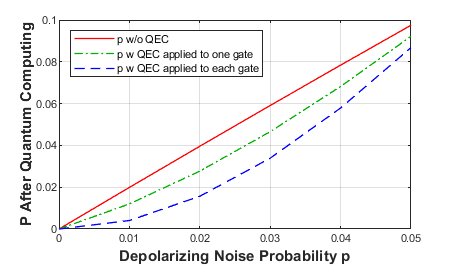}
\caption{Relationship between noise error rate and different locations of QEC in two gates case.\label{fig:twogatenoise}}
\vspace{-4mm}
\end{figure}

Figure \ref{fig:twogatenoise} shows the relationship between the depolarizing noise error rate and the locations of QEC in the case of two qubit gates. QEC operation reduces the error rate, and an increase in the number of QEC applications results in a smaller error rate.

\subsection{DP for multiple single qubit gates}
\begin{figure}[t]
\centering
\includegraphics[width=0.4\textwidth]{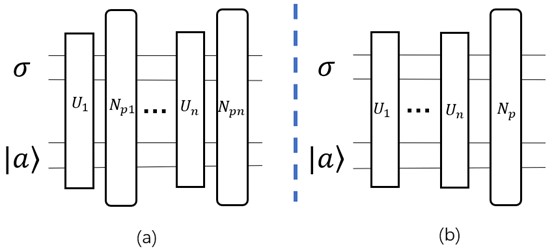}
\caption{The depolarizing noise model. (a) $\sigma$ is a quantum state, $\left | a  \right \rangle$ is auxiliary quantum bits. $U_{i}$ is arbitrary single-bit circuit gates, where $i=1...n$. $N_{pn} $ is depolarizing noise along the circuit. (b) $N_{p} $ is global depolarizing noise.}
\vspace{-4mm}
\label{fig:noise}
\end{figure}

We then extend two single qubit gates to multiple single qubit gates to obtain a generalized formula of $p'$ and the corresponding privacy budget. We assume that the noise model of the circuit is as shown in Fig. \ref{fig:noise}(a), with the depolarizing noise added to each of the $n$ single quantum gates, and the error rates are all $p$.

\textbf{No QEC operations:} When the quantum circuit consists of $n$ single quantum gates, the total error rate of the depolarizing noise is calculated in the same way as \cite{du2021quantum}. Figure~\ref{fig:noise}(b) can represent the case of Fig. \ref{fig:noise}(a), i.e., the total depolarizing noise on each gate can be expressed as a global depolarizing noise added at the end of the circuit, with a total error rate
\vspace{-1mm}
\begin{equation}
\begin{split}
    p_{3}' =1- {\textstyle \prod_{i=1}^{n}} \left ( 1-p_{i}  \right )  =1-\left ( 1-p \right ) ^{n}.
\end{split}
\end{equation}
Therefore, Eq. (\ref{oridp}) can evolve into $\epsilon=\ln_{}{[1+\frac{1-p_{3}'}{p_{3}'}dD ]}$.

\textbf{Selective QEC operations for gates:} We randomly selected $m$ gates ($m\le n$) to apply the QEC operations, and the total accuracy of these $m$ gates is $\left ( 1-p'   \right ) ^{m} $. The total accuracy of the other $n-m$ gates without QEC operation is $\left ( 1-p   \right ) ^{n-m} $. Thus, the total error rate of the circuit in this case is
\begin{equation}
\begin{split}
    p_{4}' =1- \left ( 1-p'   \right ) ^{m}\times \left ( 1-p   \right ) ^{n-m} .
\end{split}
\end{equation}
Therefore, Eq. (\ref{oridp}) can evolve into $\epsilon=\ln_{}{[1+\frac{1-p_{4}'}{p_{4}'}dD ]}$.

It should be noted that QEC applied to every gate is a special case, i.e., $m = n$, where the total error rate of the circuit with depolarizing noise is $1-\left ( 1-p'   \right ) ^{n}$ and the privacy budget is $\epsilon=\ln_{}{[1+\frac{1-[ 1-\left ( 1-p'   \right ) ^{n} ]} {1-\left ( 1-p'   \right ) ^{n}}dD ]} $.

\subsection{Multi-level concatenated Steane code}
Now we encode the logical qubit by using multi-level concatenated Steane code \cite{nikahd2017low}, which can further decrease $p$. As an example of two-level concatenated Steane codes, we encode each of the seven physical qubits using Steane code. Thus, we use $7^2=49$ physical qubits to represent a logical qubit. In a single qubit gate circuit, the total error rate of the depolarizing noise with QEC operation will become 
\begin{equation}
\begin{split}
    p_{5}' =1-\left[(1-p' )^7+7\times p' \times (1-p')^{6} \right].
\end{split}
\end{equation}
Therefore, Eq. (\ref{oridp}) can evolve into $\epsilon=\ln_{}{[1+\frac{1-p_{5}'}{p_{5}'}dD ]}$.

We find that $p_{5}'<p'$, suggesting that multi-level concatenated Steane code can have a further reduction in the depolarizing noise error rate at the cost of circuit complexity. The case of adding multi-level concatenated Steane code to multiple qubits gates will be discussed in future work.

\section{Performance evaluation\label{sec:performance}}

\subsection{Simulation setup}
In this section, we used numerical simulations to detect the relationship between the various parameters of DP formulation. Throughout the simulations, unless otherwise noted, the default settings of parameters of a single qubit gate circuit are $D = 2$, $p = 0.03$, and $d = 0.5$, where $D$ is the dimension of the Hilbert space, $p$ is the error rate of the depolarizing noise added at each gate, and $d$ is the trace distance between the two quantum states. In subsequent simulations, the remaining default parameters were kept constant as we investigated the effect of changes in individual parameters on DP.

\subsection{The impact of trace distance on QDP}
\begin{figure}[t]
\centering
\includegraphics[width=0.45\textwidth]{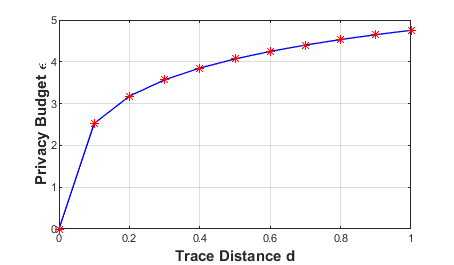}
\caption{Relationship between privacy budgets and trace distance.\label{fig:trace_distance}}
\vspace{-5mm}
\end{figure}

We investigate the effect of the trace distance between two quantum states on the privacy budget at a single qubit gate circuit. As shown in Fig. \ref{fig:trace_distance}, when the trace distance $d$ becomes larger, the privacy budget also becomes larger, which indicates that the degree of privacy protection becomes lower. The trace distance is the difference between two quantum states. When $d$ becomes larger, the difference between quantum states is larger, and therefore it becomes more difficult to achieve privacy protection. In practice, two quantum states need to be encoded, and then quantum computation is performed. The process of encoding gives the current trace distance of the two quantum states and thus the corresponding privacy budget. For two specific quantum states, the value of $d$ is usually determined, so we cannot achieve regulation of the privacy budget by adjusting $d$.

\begin{figure}[t]
\centering
\includegraphics[width=0.45\textwidth]{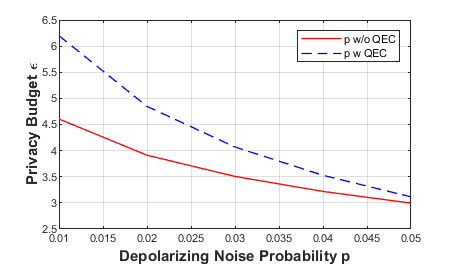}
\caption{Relationship between privacy budgets and noise error rate of one gate.\label{fig:one_gate_privacy}}
\vspace{-5mm}
\end{figure}

\begin{figure}[t]
\centering
\includegraphics[width=0.45\textwidth]{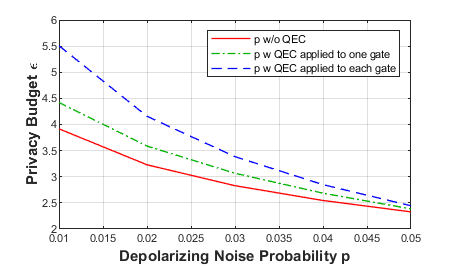}
\caption{Relationship between privacy budgets and noise error rate under different locations of QEC.\label{fig:two_gates_qecs_budget}}
\vspace{-5mm}
\end{figure}

\subsection{QEC achieves regulation of privacy budgets}

\textbf{Single qubit gate calculation:} We set the range of the depolarizing noise between 0 and 0.05. It is the threshold that guarantees the effect of QEC. Figure~\ref{fig:one_gate_privacy} shows the effect on the privacy budget brought by the variation of the noise error rate $p$. As $p$ becomes larger, the privacy budget is smaller, and the privacy protection is better, but the data usability becomes worse. Depolarizing noise in the circuit already satisfies a certain level of privacy protection. After adding QEC to the circuit, the final noise error rate $p'$ becomes smaller, so the privacy budget becomes larger. 

\textbf{Two qubit gates calculation:} The quantum circuit is two single qubit gates, with depolarizing noise added to each gate, and both with error rate $p$. We split this simulation into three cases: no QEC operation, QEC after each gate, and choosing only one quantum gate to perform QEC operation. We study its final error rate versus privacy budgets, and the results are shown in Fig. \ref{fig:two_gates_qecs_budget}. We find that the depolarizing noise error rate decreases and the privacy budget is larger than the original budget in both cases where the QEC operation is added. The case of adding QEC after each gate has a smaller error rate than selectively adding QEC. This illustrates that introducing QEC can reduce the noise and thus affect the privacy budget. Deciding whether or not to add a QEC operation to each quantum gate affects the error rate of the noise, which in turn impacts the privacy budget of QDP.

\begin{figure}[t]
\centering
\includegraphics[width=0.45\textwidth]{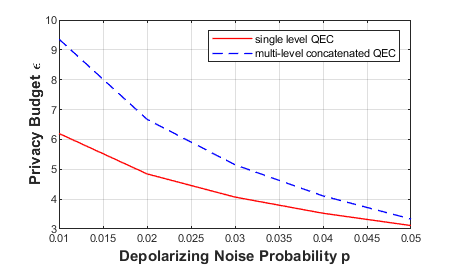}
\caption{Relationship between privacy budgets and multi-level concatenated Steane code.\label{fig:epsmulti}}
\vspace{-6mm}
\end{figure}

\textbf{Multi-level concatenated Steane code:} The privacy budget of a single qubit gate using multi-level concatenated Steane code is shown in Fig. \ref{fig:epsmulti}. Compared to a single-level Steane code, the multi-level concatenated Steane code can further diminish the noise and achieve a higher privacy budget at the cost of increased circuit complexity.

\section{Conclusion and future work}
In this paper, we use QEC method to reduce the error rate, leading to solving excessive inherent noises and gaining the target privacy budget. We find that whether to use QEC for each gate affects the depolarizing noise error rate progressively. We summarize the general noise error rate formula for applying QEC operations to multiple single qubit gates and the corresponding privacy budget $\epsilon$. In addition, we explore how multi-level concatenated Steane code can further reduce noise at the expense of circuit complexity. Our simulation results show that QEC can reduce the noise error rate and thus increase the privacy budget for a specific depolarizing noise error rate interval. Therefore, we can claim that QDP protection can be tuned/controlled through QEC operation. Since this paper mainly focuses on theoretical analysis and numerical study, we plan to conduct experiments on quantum devices to further verify the proposed ideas in the future.
\vspace{2mm}

\bibliographystyle{IEEEtran}
\bibliography{./ref}
\vspace{12pt}

\end{document}

%% file: table.tex
\begin{table*}[ht]
\caption{Error syndrome of Steane code.}
\label{table*}
\centering
\begin{tabular}{cccccccc}
\hline
\multicolumn{8}{c}{X error (M0,M1,M2=1,1,1)}                                                                                 \\ \hline
i                                     & 0        & 1        & 2        & 3         & 4         & 5         & 6          \\
Eigenvalue of (N0,N1,N2)              & (-1,1,1) & (1,-1,1) & (1,1,-1) & (1,-1,-1) & (-1,1,-1) & (-1,-1,1) & (-1,-1,-1) \\ \hline
\multicolumn{8}{c}{Z error (N0,N1,N2=1,1,1)}                                                                                 \\ \hline
i                                     & 0        & 1        & 2        & 3         & 4         & 5         & 6          \\
Eigenvalue of (M0,M1,M2)              & (-1,1,1) & (1,-1,1) & (1,1,-1) & (1,-1,-1) & (-1,1,-1) & (-1,-1,1) & (-1,-1,-1) \\ \hline
\multicolumn{8}{c}{Y error}                                                                                             \\ \hline
i                                     & 0        & 1        & 2        & 3         & 4         & 5         & 6          \\
Eigenvalue of   (M0,M1,M2)=(N0,N1,N2) & (-1,1,1) & (1,-1,1) & (1,1,-1) & (1,-1,-1) & (-1,1,-1) & (-1,-1,1) & (-1,-1,-1) \\ \hline
\end{tabular}
\end{table*}